\documentclass[prb,aps,amsmath,nofootinbib]{revtex4}

\usepackage{graphicx}% Include figure files
\usepackage{dcolumn}% Align table columns on decimal point
\usepackage{bm}% bold math

\begin{document}
\renewcommand{\thefootnote}{\fnsymbol{footnote}}

%\preprint{}

\title{Effect of different choices of the Boltzmannized flux operator 
on thermal exchange and recombination reactions}


\author{Koichi Saito}
\email[]{ksaito@tohoku-pharm.ac.jp}
%\thanks{}
\affiliation{Tohoku Pharmaceutical University, Sendai 981-8558, Japan}


\date{\today}

\begin{abstract}
The rate constants for recombination and exchange reactions are
calculated using the flux correlation approach with 
a general form of the Boltzmannized flux operator, which 
can simultaneously describe the Kubo and traditional 
{\em half-split} forms.  First, we consider an exactly solvable model, i.e., 
the free particle case, in terms of a new scaling function. 
Next, as a non-trivial case, we study the recombination and exchange 
reactions at very high pressure. 
Since the rate constant is calculated by Laplace transform of the
flux correlation function, the result depends on how 
the Boltzmannized flux operator is chosen. 
We find that a choice of the flux operator affects the rate constant 
considerably. For the 
recombination reaction, the ratio of the rate 
constant in the {\em half-split} form to that in the 
Kubo form approaches zero in the high pressure limit. 
\end{abstract}

%\pacs{}
%\keywords{}

\maketitle


%
\section{\label{sec:1}Introduction}
\setcounter{footnote}{1}

To study a chemical reaction at the most detailed level, it is 
necessary to treat the action of the time evolution operator, 
$\exp(-i\hat{H}t)$ ($\hat{H}$ is the Hamiltonian of the system),
\footnote{We use the natural unit, i.e., $\hbar = c = 1$.} 
onto a (given) initial wave function or density matrix, which is
a function of total energy $E$ and total angular momentum $J$.
Such quantum scattering calculations have actually been studied for 
simple chemical reactions.\cite{cal} A 
time-dependent scattering formalism based on the $S$-matrix Kohn
variational approach\cite{kohn} or a coupled channel method in
hyperspherical coordinates\cite{hyper} has usually been used to evaluate 
the quantum reactive scattering cross sections.  The number of open vibration-rotation channels, however, increases very rapidly as 
thermally accessible collision energy becomes high.  Therefore, at 
high energy the exact quantum state-to-state calculations would not 
be feasible even for a simple reaction.  Furthermore, if it is {\em only} 
the rate constant that is desired, such a complete calculation for 
all state-to-state information is {\em not} economical. 

A totally different approach for treating chemical reactions is that 
based on a correlation between quantum flux operators in the statistical 
thermodynamics. In the early 60's, Yamamoto\cite{yama} first formulated 
an exact expression for the rate constant as an application of the general
statistical mechanical theory of irreversible process, which was
established by Kubo et al.\cite{kubo,kubo2} and Mori.\cite{mori} 
(See also Ref.\onlinecite{wol}.)  Independently, 
Miller et al.\cite{miller,miller2,miller3} have developed a method for the
rate constant using a time integral of the flux-flux autocorrelation function,
which is also exact in the limit that the dynamics is extended to $t \to \infty$.
The feasibility of this approach depends on how to evaluate the time evolution operator for the system, and the correlation is usually calculated using 
the Feynman path integral technique.\cite{path,path2}
Although this method is powerful and convenient to obtain the rate constant, 
the direct application to large, complicated systems is still difficult 
because of the notorious sign problem.\cite{sign,sign2}  
One approach to improve this situation is a filtering or smoothing 
technique\cite{f1} such as the stationary-phase Monte Carlo.\cite{spmc}  
Recently, semiclassical (SC) approaches, implemented via the initial-value representation (SC-IVR), have received a rebirth of interest. 
(Van Vleck\cite{van} first discussed a 
drastic approximation based on the semiclassical picture.) 
A number of studies\cite{yama2,yi,yama3} have been carried out along SC-IVR 
and have demonstrated the 
capability of these approaches to describe various quantum effects in chemical 
reactions. 

It has also been shown how a quantum mechanical version of the
Lindemann mechanism for collisional recombination, where 
the process is affected by the bath gas $M$, 
\begin{eqnarray}
A + B &\rightleftharpoons& AB^* , \label{ab1} \\
AB^* + M &\to& AB + M , \label{ab2}
\end{eqnarray}
can be handled by the flux-flux autocorrelation function for
the $A-B$ collision.\cite{lind,lind2,appl}  Some 
applications of this theory are listed in Refs.\onlinecite{appl2,appl3}. 
It is furthermore possible to generalize the formalism to include
chemical reactions as well as recombination:
\begin{eqnarray}
A + BC &\rightleftharpoons& ABC^* \to AB + C , \label{abc1} \\
ABC^* + M &\to& ABC + M . \label{abc2}
\end{eqnarray}
Equations (\ref{abc1}) and (\ref{abc2}) simultaneously describe 
the recombination
process ($A + BC \to ABC$) and the exchange reaction ($A + BC \to AB +
C$).\cite{lind2} This method has been applied to the interesting
(combustion) reactions ($O + OH \rightleftharpoons H + O_2$) and the
recombination reactions ($O + OH + M \to HO_2 + M \gets H + O_2 + M$), 
which are very important in atmospheric chemistry.\cite{atm,atm2} 

As mentioned above, there are two different ways to calculate the rate 
constant for chemical reactions, i.e., the Yamamoto approach, in which 
the Kubo form of the flux operator is used, and the Miller approach, in 
which the traditional {\em half-split} form is chosen. The two approaches can 
provide the same result to the rate constant because it is given in terms of the integral of the flux-flux correlation function 
with respect to time.\cite{miller,miller3}  
However, the shapes of the correlation functions calculated by the 
two approaches are quite different from each other. Therefore, for 
the recombination 
and exchange reactions (like Eqs.(\ref{abc1}) and (\ref{abc2})) 
the rate constants in the two approaches could be 
different, because they are given 
by Laplace transforms of the flux-flux correlation functions.\cite{saito} 
The purpose of this paper is to demonstrate the effect of different 
choices of the Boltzmannized flux operator 
on the rate constants for recombination and exchange reactions. 

{}First, we summarize the correlation function method briefly 
in Sec.\ref{sec:2}.   
The expressions of the correlation functions in the Yamamoto and 
Miller approaches can be unified using 
a general form of the Boltzmannized flux 
operator. The difference between the two approaches is discussed 
explicitly.  The rate constants for recombination and exchange
processes are also studied.  In Sec.\ref{sec:3}, we consider an exactly
solvable model, i.e., the free particle case, in terms of 
a new scaling function. In Sec.\ref{sec:4}, as a non-trivial case, 
we study the recombination and exchange reactions at very high 
pressure. Finally, the summary and conclusion are given in Sec.\ref{sec:5}.

\section{\label{sec:2}Flux-flux correlation approach to rate constants}

\begin{figure*}
\includegraphics[height=12cm]{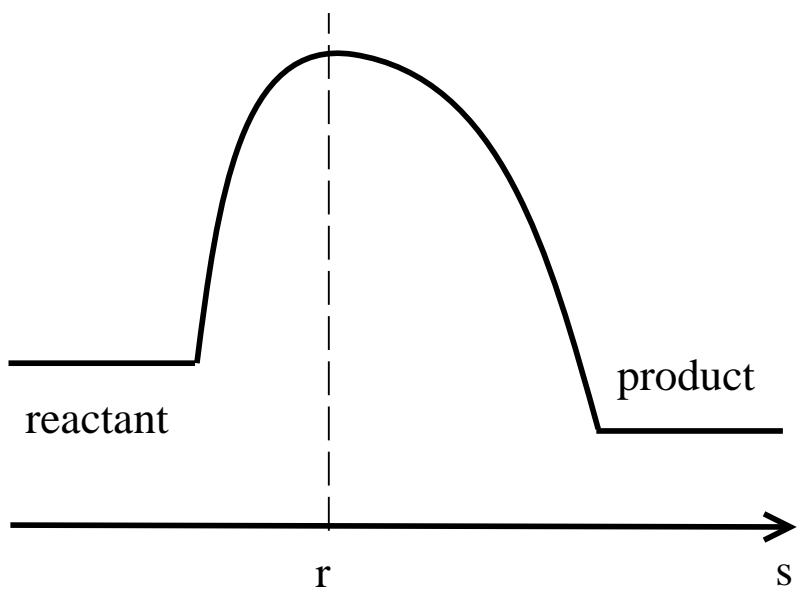}%
\caption{\label{f:reac} Sketch of a potential surface in one-dimensional
reaction versus the reaction coordinate $s$. The dividing point is 
denoted by $r$.}
\end{figure*}

The quantum mechanically exact expression for a thermal 
rate constant $k(T)$ can be written in terms of 
the flux correlation function\cite{miller}\footnote{For simplicity, 
we consider a reaction in one-dimension.} 
\begin{equation}
k(T) = Q_0(T)^{-1} \lim_{t\to\infty} C_{s}(r;t) ,  \label{kt}
\end{equation}
where $Q_0(T)$ is the reactant partition function per unit volume and 
$C_s(r;t)$ is the {\em flux-side} correlation function defined by
\begin{equation}
C_{s}(r;t) = {\rm tr}[e^{-\beta\hat{H}}\hat{F}(r)\hat{{\cal P}}] .  \label{cfs}
\end{equation}
Here, $\hat{F}(r)$ is the {\it bare} flux operator given by 
\begin{equation}
\hat{F}(r) = i [\hat{H}, h(\hat{s})]_{s=r} ,   \label{fff}
\end{equation}
with $h$ the Heaviside step function and $s$ the reaction coordinate -- 
see Fig.\ref{f:reac}. $h(s)$ takes the value of $0(1)$ in the 
reactant (product) side of the dividing point $r$.  
In Eq.(\ref{cfs}), $\hat{{\cal P}}$ is the projection operator defined by
\begin{equation}
\hat{{\cal P}} = e^{i\hat{H}t}h(\hat{p})e^{-i\hat{H}t} ,  \label{proj}
\end{equation}
with $\hat{p}$ the momentum operator. 

{}For the rate constant, two different approaches have been proposed so far: 
one is the Yamamoto approach,\cite{yama} which is based on the linear 
response theory (or the so-called Kubo formula),\cite{kubo,kubo2} and the 
other is the flux-flux autocorrelation function method 
developed by Miller et al.\cite{miller,miller2,miller3} 
In the Miller approach, Eq.(\ref{cfs}) are modified by following two useful 
facts: the first one is made by noting that $\hat{{\cal P}}$ and $\hat{H}$ 
commute each other (i.e., $[\hat{{\cal P}}, \hat{H}] = 0$) in the limit $t\to\infty$, and the second is to replace 
$h(\hat{p})$ by $h(\hat{s})$ in Eq.(\ref{proj}), which is also 
correct in the limit $t\to\infty$. By virtue of these modifications, 
the flux-side correlation function in the Miller approach 
can be rewritten as  
\begin{equation}
C_{s}^M(r;t) = {\rm tr}[\hat{F}^M(\beta,r)e^{i\hat{H}t}
h(\hat{r})e^{-i\hat{H}t}] , 
\label{mcfs}
\end{equation}
where the superscript $M$ stands for "Miller".  Here,  
$\hat{F}^M(\beta,r)$ is the {\em half-split} Boltzmannized flux operator
\begin{equation}
\hat{F}^M(\beta,r) = e^{-\beta\hat{H}/2}\hat{F}(r)e^{-\beta\hat{H}/2} . 
\label{mflux}
\end{equation}
Using Eq.(\ref{fff}), 
the rate constant can also be expressed in terms of the {\em flux-flux} 
correlation function $C(r,r;t)$ as
\begin{equation}
Q_0(T) k(T) = \int_0^{\infty} dt \ C^M(r,r;t) ,  \label{kt2}
\end{equation}
where
\begin{equation}
C^M(r,r;t) = {\rm tr}[\hat{F}^M(\beta,r)e^{i\hat{H}t}\hat{F}(r)
e^{-i\hat{H}t}] . 
\label{mcff}
\end{equation}

On the other hand, the linear response theory gives the rate 
constant 
\begin{equation}
Q_0(T) k(T) = \int_0^{\infty} dt \ C^Y(r,r;t) ,  \label{kt3}
\end{equation}
where 
\begin{equation}
C^Y(r,r;t) = {\rm tr}[\hat{F}^Y(\beta,r)e^{i\hat{H}t}\hat{F}(r)
e^{-i\hat{H}t}] , 
\label{ycff}
\end{equation}
with 
\begin{equation}
\hat{F}^Y(\beta,r) = \frac{1}{\beta} \int_0^\beta d\lambda \ 
e^{-(\beta-\lambda)\hat{H}}\hat{F}(r)e^{-\lambda\hat{H}} .  
\label{yflux}
\end{equation}
Here, the superscript $Y$ stands for "Yamamoto". 

If a general form of the Boltzmannized flux operator is introduced 
by\cite{yama3}
\begin{equation}
\hat{F}^\kappa(\beta,r) = \frac{1}{\kappa\beta} 
\int_{(1-\kappa)\beta/2}^{(1+\kappa)\beta/2} d\lambda \ 
e^{-(\beta-\lambda)\hat{H}}\hat{F}(r)e^{-\lambda\hat{H}} ,   
\label{gflux}
\end{equation}
where $\kappa$ is a parameter ($0\leq \kappa \leq 1$), 
the correlation functions in the Miller 
and Yamamoto approaches can be unified as
\begin{equation}
C^\kappa(r,r;t) = {\rm tr}[\hat{F}^\kappa(\beta,r)e^{i\hat{H}t}
\hat{F}(r)e^{-i\hat{H}t}] . 
\label{gcff}
\end{equation}
It is easy to check that in the limit $\kappa \to 0(1)$ Eq.(\ref{gcff}) 
reproduces the correlation function in the Miller (Yamamoto) approach. 

Using Eq.(\ref{fff}), the integral with respect to 
$\lambda$ in the general form of the Boltzmannized flux operator 
can be performed:
\begin{equation}
\hat{F}^\kappa(\beta,r) = \frac{i}{\kappa\beta} \left[
e^{-(1-\kappa)\beta\hat{H}/2}h(\hat{r})e^{-(1+\kappa)\beta\hat{H}/2} 
- e^{-(1+\kappa)\beta\hat{H}/2}h(\hat{r})e^{-(1-\kappa)\beta\hat{H}/2}
\right] .    
\label{gflux2}
\end{equation}
Combining the {\em partial} Boltzmann operator, 
$e^{-(1\pm\kappa)\beta\hat{H}/2}$, 
with the real-time evolution operator $e^{\pm i\hat{H}t}$, we obtain 
two complex-time evolution operators: $e^{-i\hat{H}t_+}$ and 
$e^{-i\hat{H}t_-}$ with $t_\pm = t - i (1\pm\kappa)\beta/2$. 
Then, the correlation function is rewritten as
\begin{equation}
C^\kappa(r,r;t) = 
\frac{2}{\kappa\beta} \Im \ {\rm tr}[\hat{F}(r)e^{i\hat{H}t_+^*}
h(\hat{r})e^{-i\hat{H}t_-}] , 
\label{gcff1}
\end{equation}
where $\Im$ stands for taking the imaginary part. 
Performing the trace operation in Eq.(\ref{gcff1}), 
the correlation function reads
\begin{eqnarray}
C^\kappa(r,r';t) &\equiv& \frac{2}{\kappa\beta} \Im \ {\rm tr} 
[\hat{F}(r)e^{i\hat{H}t_+^*}
h(\hat{r'})e^{-i\hat{H}t_-}]  \nonumber \\
&=& \frac{1}{m\kappa\beta} \Im \int_{r}^\infty ds
\left[ i \langle s | e^{-i\hat{H}t_+} | s' \rangle^*
\frac{\partial}{\partial s'}
\langle s | e^{-i\hat{H}t_-} | s' \rangle \right. \nonumber \\
&-& \left. i \langle s | e^{-i\hat{H}t_-} | s' \rangle
\frac{\partial}{\partial s'} \langle s | e^{-i\hat{H}t_+} | 
s' \rangle^* \right]_{s'= r'} , 
\label{gcff2}
\end{eqnarray}
where $\langle s | e^{-i\hat{H}t} | s' \rangle$ is the propagator 
in the coordinate representation. 
The correlation function $C^\kappa(r,r;t)$ is given in 
the limit $r' \to r$. 

\begin{figure*}
\includegraphics[height=12cm]{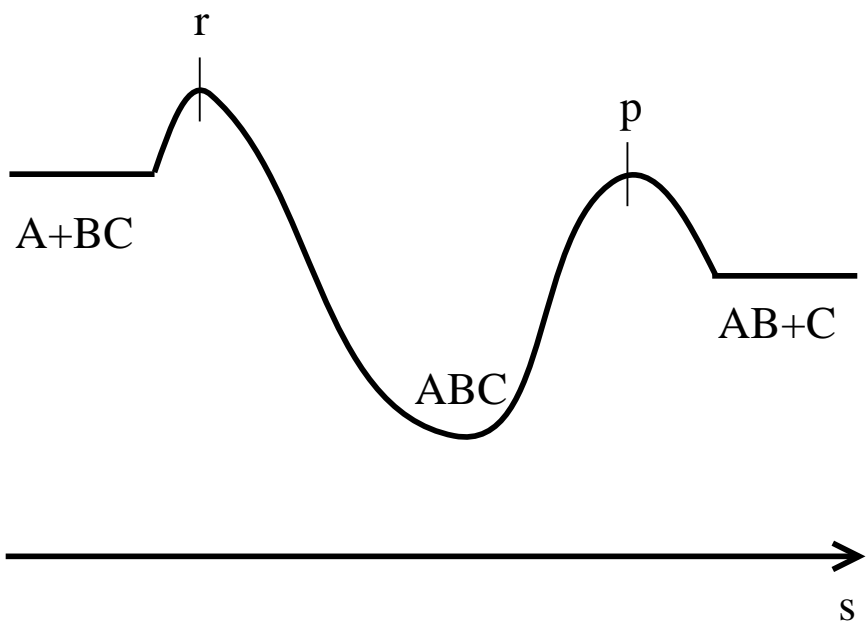}%
\caption{\label{f:diag} One-dimensional schematic diagram of the 
potential for $A+BC \to AB+C$ reaction.
The compound region ($ABC$) is bounded by the dividing points on
reactant ($r$) and product ($p$) sides.}
\end{figure*}

It is possible to generalize the flux correlation approach to treat  
recombination and exchange reactions.\cite{appl,lind,lind2,appl2,appl3} 
Let us consider the reaction of $A + BC \to AB + C$ and $ABC$ 
in one-dimension (see Eqs.(\ref{abc1}) and (\ref{abc2}) 
and Fig.\ref{f:diag}).  The rate constants for the
exchange ($A + BC \to AB + C$) and recombination ($A + BC \to ABC$)
reactions are again given by the average of the
flux $\hat{F}(r)$ and the projection operator  
${\hat {\cal P}}$ over the Boltzmann distribution, where $\hat{F}(r)$ 
describes the flux at the reactant dividing point $r$ (see 
Fig.\ref{f:diag}). Similarly, we can define the flux operator 
$\hat{F}(p)$ at the product dividing point $p$. 

Because the probability of the system experiencing a deactivating
($ABC^* + M \to ABC + M$) collision with the bath gas $M$ can be evaluated
by $1-e^{- \eta t}$ at time $t$ ($\eta$ describes the frequency of
deactivating collisions and it depends on pressure $P$ and $T$ of the bath
gas), the recombination probability is proportional to 
$1 - e^{- \eta \tau}$, 
where $\tau$ is the time the trajectory (it is on $r$ at $t=0$) is in
the compound region (see Fig.\ref{f:diag}).  On the other hand, for 
the exchange reaction, the probability is given by $e^{- \eta \tau_p}$, where $\tau_p$ is the time the trajectory exists through the point $p$.  

Using these probabilities and the correlation function, 
the rate constants for the recombination and exchange
reactions are finally given by\cite{lind2}
\begin{eqnarray}
Q_0(T) k_{rec}^\kappa(T,P) &=& \int_0^\infty dt \ e^{- \eta t}
[C^\kappa(r,r;t) - C^\kappa(r,p;t)] , \label{krec}  \\
Q_0(T) k_{exc}^\kappa(T,P) &=& \int_0^\infty dt \ e^{- \eta t} 
C^\kappa(r,p;t) , 
\label{kexc}
\end{eqnarray}
where the relation 
\begin{equation}
\int_0^\infty dt \ C^\kappa(r,r;t) = 
\int_0^\infty dt \ C^\kappa(r,p;t) 
\label{crel}
\end{equation}
holds. This ensures that the recombination rate
vanishes in the limit $\eta \to 0$. 

Because the rate constants for recombination and exchange 
processes are calculated by the Laplace transforms of the
correlation functions, it is clear that the evaluated results 
depend on $\kappa$, that is, the shape of the correlation function 
affects the rate constants.\cite{saito}  
Of course, the integral of the correlation 
function with respect to time itself does not depend on  
$\kappa$.  

\section{\label{sec:3}Solvable model -- free particle case}

In this section and the next section, we calculate the rate constants for
recombination and exchange reactions using the Feynman path integral
technique.\cite{path,path2} 
A huge calculation is usually required to obtain the exact
matrix elements of propagators for a realistic system. Furthermore, it 
is necessary to take some approximations 
and numerical techniques to perform 
it.\cite{spmc,yama2,yi,yama3,numeric,numeric2,numeric3}  
Because the aim of this paper is to show how the rate constant for 
recombination or exchange reaction depends 
on $\kappa$ (i.e., a choice of the Boltzmannized flux operator), it would 
be more intuitive and useful to first consider a simple system rather 
than 
a complicated case.  We here study an analytically solvable model, 
i.e., the free particle case in one-dimension, and discuss a non-trivial 
case in the next section. 

\begin{figure*}
\includegraphics[height=12cm]{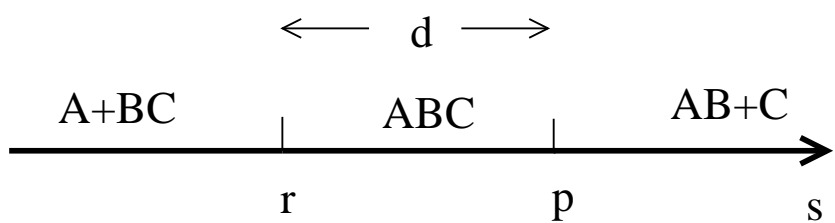}%
\caption{\label{f:free} Same as Fig.\protect\ref{f:diag}, but for 
the free particle case.}
\end{figure*}

If no potential acts on the system (see Fig.\ref{f:free}), 
the propagator in the coordinate representation is 
easily calculated by the path integral.\cite{path} 
The flux-flux correlation function Eq.(\ref{gcff2}) 
then gives
\begin{eqnarray}
C_0^\kappa(r,p;t) &=& \frac{1}{4\pi\kappa\beta^2\sqrt{D_+D_-}}
\exp\left[ -\frac{m\beta d^2(t^2+(1-\kappa^2)\beta^2/4)}{2D_+D_-} 
\right]  \nonumber \\
&\times& 
\left[ (2tA - \kappa\beta B) \sin X + (2t B + \kappa\beta A) \cos X 
\right] , \label{fckrp}
\end{eqnarray}
where $D_\pm = t^2 + (1\pm\kappa)^2\beta^2/4$, 
$X = \kappa m t \beta^2 d^2 / 2D_+D_-$ and 
\begin{eqnarray}
A &=& \left[ \left( \sqrt{D_+}+t \right) \left( \sqrt{D_-}+t \right) 
\right]^{1/2} + 
\left[ \left( \sqrt{D_+}-t \right) \left( \sqrt{D_-}-t \right) 
\right]^{1/2} , \label{aa} \\
B &=& \left[ \left( \sqrt{D_+}+t \right) \left( \sqrt{D_-}-t \right) 
\right]^{1/2} - 
\left[ \left( \sqrt{D_+}-t \right) \left( \sqrt{D_-}+t \right) 
\right]^{1/2} , \label{bb}
\end{eqnarray}
with $d$ the distance between $r$ and $p$ and $m$ the reduced mass. 
Note that, as it should be, the correlation depends on {\em only}
the distance $d$ and is independent of the positions $r$ and $p$. 

If $\kappa = 1$ (the Yamamoto approach), Eq.(\ref{fckrp}) gives 
\begin{eqnarray}
C_0^Y(r,p;t) &=& \frac{1}{2\pi\beta^2\sqrt{2t(t^2+\beta^2)}}
\exp\left[ - \frac{m \beta d^2}{2(t^2+\beta^2)} \right] \nonumber \\
&\times& \left[ (\sqrt{t^2+\beta^2} + t)^{3/2} \sin X' +
(\sqrt{t^2+\beta^2} - t)^{3/2} \cos X' \right] , \label{fcyrp}
\end{eqnarray}
where $X' = m \beta^2 d^2 / 2t(t^2+\beta^2)$, and 
\begin{equation}
C_0^Y(r,r;t) = \frac{(\sqrt{t^2+\beta^2} - t)^{3/2}}
{2\pi\beta^2\sqrt{2t}(t^2+\beta^2)^{1/2}} . \label{fcyrr}
\end{equation}
On the other hand, taking the limit $\kappa \to 0$ in Eq.(\ref{fckrp}), 
we obtain the correlation functions in the Miller approach: 
\begin{equation}
C_0^M(r,p;t) = \frac{1}{4\pi(t^2+\beta^2/4)^{3/2}}
\left[ \frac{\beta}{2} + \frac{2mt^2d^2}{t^2+\beta^2/4} \right]
\exp\left[ - \frac{m \beta d^2}{2(t^2+\beta^2/4)} \right] , \label{fcmrp}
\end{equation}
and
\begin{equation}
C_0^M(r,r;t) = \frac{\beta}{8\pi(t^2+\beta^2/4)^{3/2}} . \label{fcmrr}
\end{equation}

Now it is very convenient to introduce a new scaling function 
\begin{eqnarray}
S_{rp}^\kappa(x,c) &\equiv& 2\pi\beta^2 C_0^\kappa(r,p;t) \nonumber \\
&=& \frac{1}{2 \kappa \sqrt{{\bar D}_+{\bar D}_-}}
\exp\left[ - c \left(\frac{x^2+(1-\kappa^2)/4)}{{\bar D}_+{\bar D}_-} 
\right) \right] 
\nonumber \\
&\times& \left[ (2x{\bar A} - \kappa {\bar B}) \sin X + 
(2x {\bar B} + \kappa {\bar A}) \cos X 
\right] , \label{sc1}
\end{eqnarray}
where $x (= t/\beta)$ and $c (=md^2/2\beta)$ are {\em dimensionless} 
variables. Since the scaling function itself is a function of $x$ and $c$, 
it is also dimensionless. 
Here, ${\bar D}_\pm = x^2 + (1\pm\kappa)^2/4$, 
$X = \kappa c x / {\bar D}_+{\bar D}_-$, and 
${\bar A}$ and ${\bar B}$ are given by replacing $D_\pm$ with 
${\bar D}_\pm$ and $t \to x$ in Eqs.(\ref{aa}) and (\ref{bb}), respectively. 
If we set $r=p$ (or $d=0$), we obtain 
\begin{equation}
S_{rr}^\kappa(x) \equiv 2\pi\beta^2 C_0^\kappa(r,r;t)
= \frac{1}{2 \kappa \sqrt{{\bar D}_+{\bar D}_-}}
(2x {\bar B} + \kappa {\bar A}) , \label{sc2}
\end{equation}
which is a function of {\em only} $x$ because $c=0$. 

\begin{figure*}
\includegraphics[height=11cm]{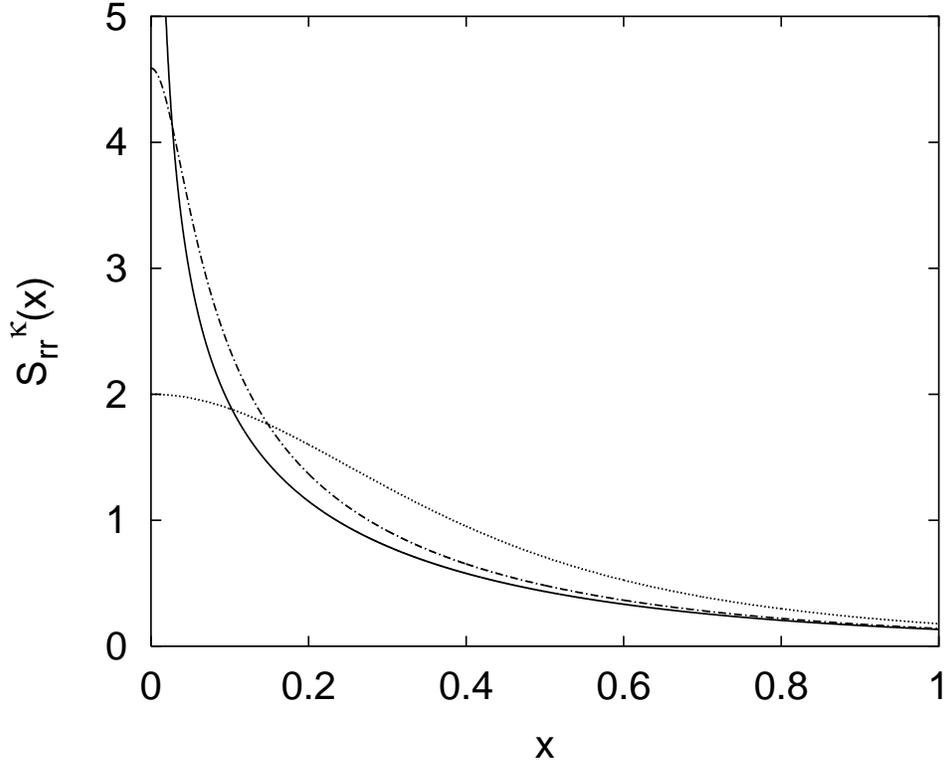}%
\caption{\label{f:srr} Scaling function $S_{rr}^\kappa(x)$ for 
the free particle case.
The dotted (dot-dashed) [solid] curve is for $\kappa = 0 (0.9) [1]$.}
\end{figure*}

\begin{figure*}
\includegraphics[height=11cm]{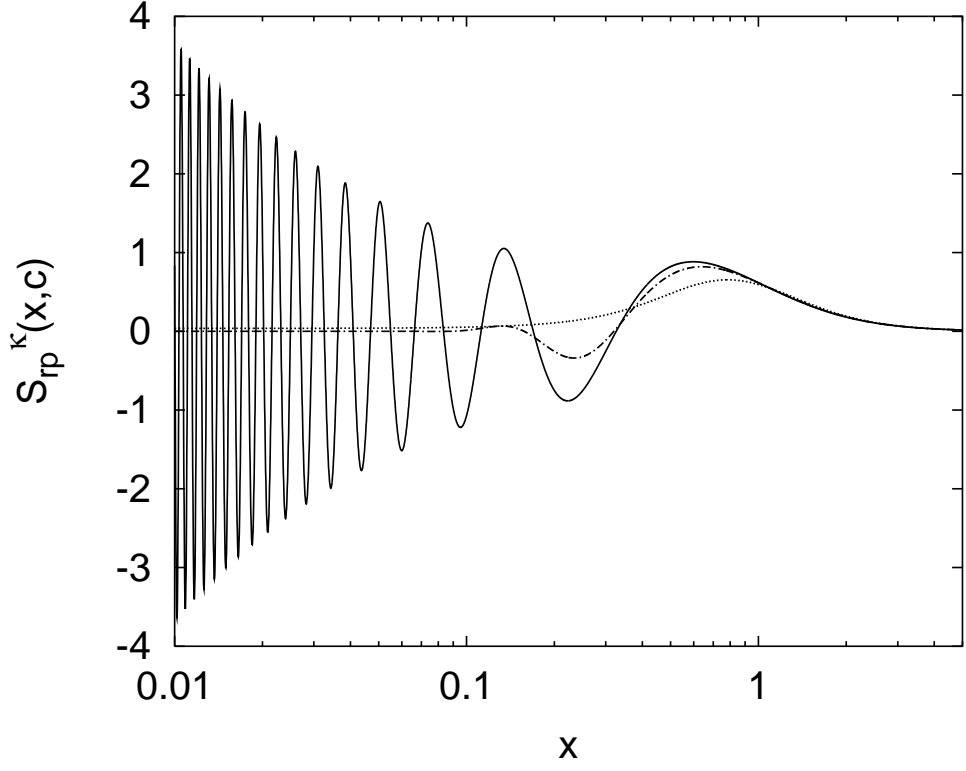}%
\caption{\label{f:srp} Scaling function $S_{rp}^\kappa(x,c)$ for 
the free particle case.
The dotted (dot-dashed) [solid] curve is for $\kappa = 0 (0.9) [1]$. 
We take $c = 1.0$.}
\end{figure*}

We show these scaling functions in Figs.\ref{f:srr} and \ref{f:srp}. 
Note that the integral of the scaling function with respect 
to $x$ does not, of course, depend on $\kappa$. 
In Fig.\ref{f:srr} the scaling function for $\kappa=1$ 
(Yamamoto approach) is divergent like $\sim 1/\sqrt{x}$ at $x=0$. 
Taking the limit $x \to 0$  
implies that $t$ approaches $0$ but $\beta$ is finite or 
$T$ is very low ($\beta \to \infty$) but $t$ is finite.  
In the Miller approach ($\kappa=0$), 
the scaling function is finite at $x=0$: $S_{rr}^{\kappa=0}(0)=2$.  For 
arbitrary $\kappa$, we find $S_{rr}^\kappa(0)=2/\sqrt{1-\kappa^2}$. 
By contrast, in the limit $x \to \infty$ (which corresponds to the case 
where $T$ is very high ($\beta \to 0$) but $t$ is finite or 
$t \to \infty$ but $\beta$ is finite), 
the scaling function does not depend on $\kappa$ and 
behaves like $\sim 1/4x^3$. 

In Fig.\ref{f:srp} we take $c = 1.0$ to illustrate 
the scaling function $S_{rp}^\kappa$ clearly.
It is remarkable that the interference
effect in the correlation is fully taken into account in the
Yamamoto approach ($\kappa = 1$).  Hence the scaling 
function oscillates very rapidly near the origin. 
Contrastingly, in the case of $\kappa \ne 1$ the interference is
averaged and the shape is very smooth.  The value of the scaling 
function at $x=0$ is given by $(2/\sqrt{1-\kappa^2})\exp(-4c/(1-
\kappa^2))$ for $\kappa \neq 1$. The {\em partial-split} form of the 
Boltzmannized flux operator (i.e., $0 \le \kappa < 1$) is thus 
very convenient for 
actual computations. For large $x$ the scaling function is again 
independent of $\kappa$ and behaves like $\sim (1+8c)/4x^3$. 

Next, we study the rate constants for recombination and exchange
reactions, which are given by Eqs.(\ref{krec}) and (\ref{kexc}), 
respectively.  Using the scaling functions, we first calculate 
the $(rr)$- and $(rp)$-rate constants, which are, respectively, defined by
\begin{eqnarray}
Q_0(T) k_{rr}^\kappa &=& \frac{1}{2\pi\beta}
\int_0^\infty dx \ e^{-\alpha x} S_{rr}^\kappa(x) , \label{fkrr}  \\
Q_0(T) k_{rp}^\kappa &=& \frac{1}{2\pi\beta}
\int_0^\infty dx \ e^{-\alpha x} S_{rp}^\kappa(x,c) , \label{fkrp}
\end{eqnarray}
with $\alpha (= \beta \eta)$ a dimensionless variable. 
Then, the rate constants are given as $k_{rec}^\kappa = k_{rr}^\kappa -
k_{rp}^\kappa$ and $k_{exc}^\kappa = k_{rp}^\kappa$.

In order to convert the collision frequency $\eta$ to more familiar
variables, we approximate the collisional deactivation rate constant by
an expression given by the hard sphere collision
theory. Furthermore, if one uses the ideal gas
expansion, the frequency can be expressed by\cite{lind2}
\begin{equation}
\eta = k_{deact}[M] = P \sqrt{\frac{2000}{T}} \times 10^{-11} ,
\label{eta}
\end{equation}
with $\eta$ in fs$^{-1}$, $P$ in Pa and $T$ in K. Then, we find
\begin{equation}
\alpha = \beta \eta \simeq 3.42 \times \frac{P}{T^{3/2}} \times
10^{-6} .
\label{alpha}
\end{equation}
The factor $c$ is also converted as
\begin{equation}
c = \frac{md^2}{2 \beta} = 0.0103 \times A_mTd^2 ,
\label{c1}
\end{equation}
with $A_m$ the reduced mass of the system in atomic mass units 
and $d$ in {\AA}.

\begin{figure*}
\includegraphics[height=11cm]{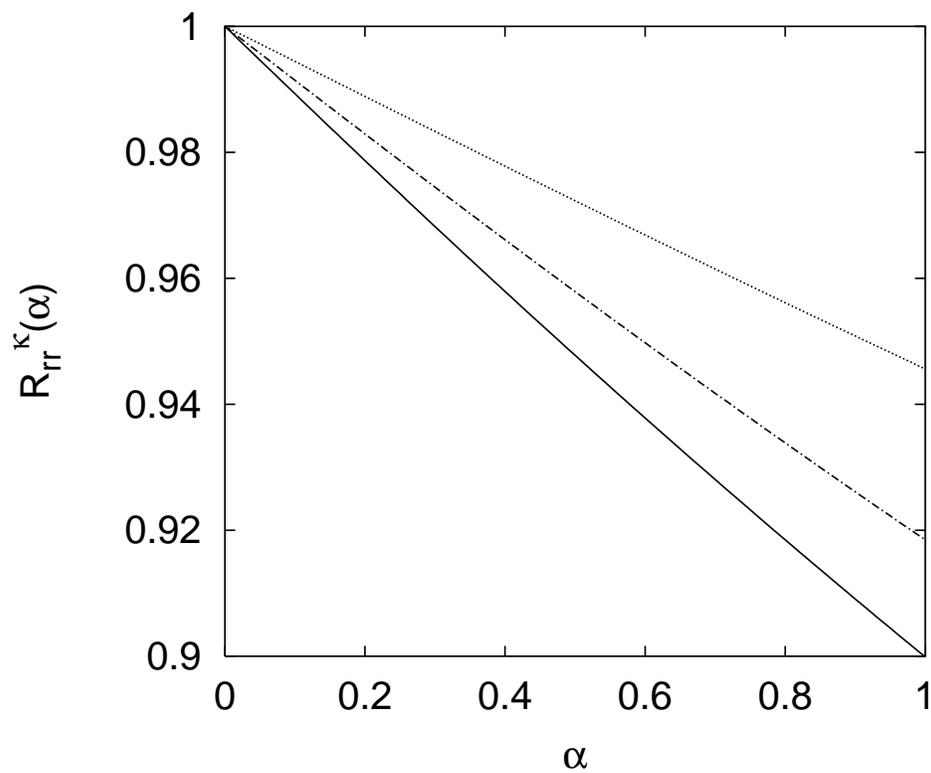}%
\caption{\label{f:Rrr} Ratio of the $(rr)$-rate constants in 
the free particle case. 
The solid (dot-dashed) [dotted] curve is for $\kappa = 0 (0.5) [0.75]$.}
\end{figure*}

\begin{figure*}
\includegraphics[height=11cm]{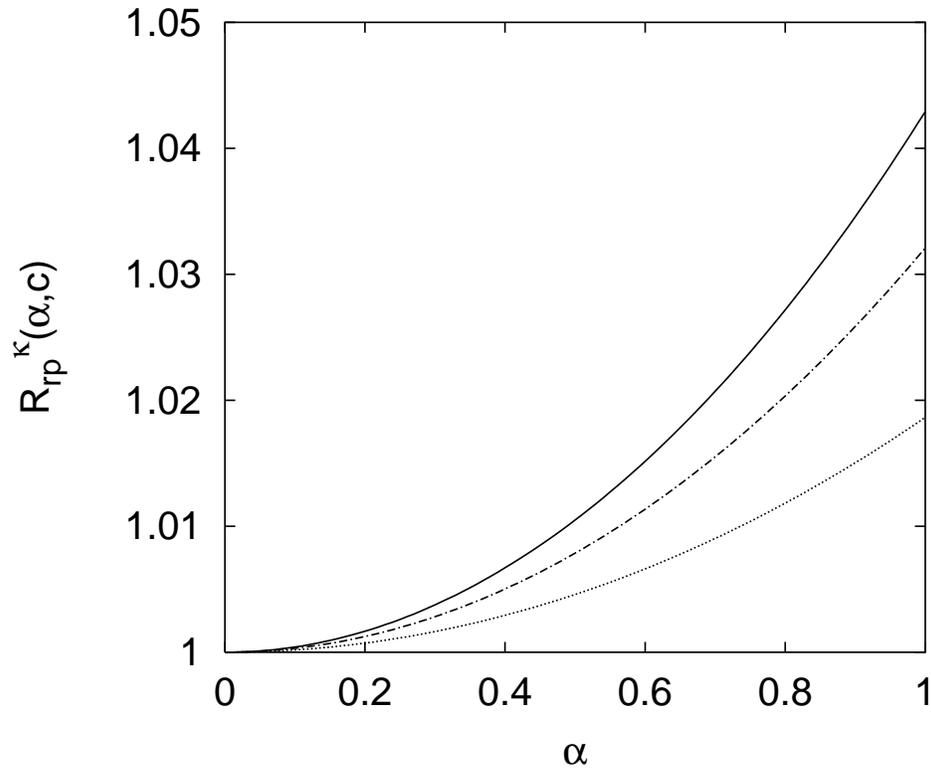}%
\caption{\label{f:Rrp} Same as Fig.\protect\ref{f:Rrr}, but for 
the $(rp)$-rate constant. 
The solid (dot-dashed) [dotted] curve is for $\kappa = 0 (0.5) [0.75]$. 
We take $c = 20.0$.}
\end{figure*}

Now we are in a position to show our results for the free particle 
case.  We define the ratio of the rate constant for 
arbitrary $\kappa$ to that for $\kappa=1$ (Yamamoto approach): 
\begin{eqnarray}
R_{rr}^\kappa(\alpha) = k_{rr}^\kappa/k_{rr}^{\kappa=1} , 
\label{fRrr} \\
R_{rp}^\kappa(\alpha,c) = k_{rp}^\kappa/k_{rp}^{\kappa=1} . 
\label{fRrp}
\end{eqnarray}
The ratio $R_{rr}^\kappa$ is a function of {\em only} $\alpha$, 
through which it depends on $T$ and $P$, while the ratio  
$R_{rp}^\kappa$ depends on $\alpha$ and $c$. 
In Figs.\ref{f:Rrr} and \ref{f:Rrp}, we illustrate the two ratios for 
the free particle case. Here, we choose $0 \le \alpha \le 1$, which, for 
example, covers the region of $P \alt 1.5$ GPa at $T \sim 300$ K.
In Fig.\ref{f:Rrp}, $c$ is fixed to be $20$ because the ratio is not 
sensitive to it in the region of $0 \le \alpha \le 1$. 
In $R_{rr}^\kappa$, the ratio decreases linearly and it is about $0.9$ 
($10\%$ reduction) at $\alpha = 1$ in the case of $\kappa=0$. 
Note that a large value of $\alpha$ corresponds 
to the case of high $P$ and low $T$, and that the ratio at large 
$\alpha$ is mainly 
determined by the correlation function (or the scaling function) at 
$x \simeq 0$ (see Fig.\ref{f:srr}).  By contrast, in $R_{rp}^\kappa$ the 
deviation of the ratio from unity is not large, and the ratio for 
$\kappa = 0$ is enhanced by about $4-5$\% at $\alpha =1$. 

Combining the $(rr)$- and $(rp)$-rate constants, we can calculate the
ratio of the recombination rate constants, $R_{rec}^\kappa =
k_{rec}^\kappa/k_{rec}^{\kappa=1}$.  Because the $(rp)$-rate constant 
itself is, however, much smaller than the $(rr)$-rate constant, 
the ratio $R_{rec}^\kappa$ is very close to $R_{rr}^\kappa$.  We find that 
$R_{rec}^{\kappa=0}$ is again about $0.9$ at $\alpha = 1$.  On the 
other hand, the ratio for the exchange process is given by 
$R_{exc}^\kappa =
k_{exc}^\kappa/k_{exc}^{\kappa=1} = R_{rp}^\kappa$. 

\section{\label{sec:4}Rate constants at high pressure limit}

As a non-trivial example, we consider the rate constants for 
recombination and exchange reactions at very high pressure. 
As we have seen in the previous section, it is very notable that at 
short time (or small $x$) the 
correlation function (or the scaling function) in the Yamamoto 
approach is quite different from the other case ($\kappa \neq 1$). The 
rate constant is calculated by the Laplace transform of the 
correlation function (see Eqs.(\ref{krec}) and (\ref{kexc})), in which the integrand involves the factor $\exp(-\eta t)$ and  
$\eta$ is proportional to pressure.  Thus, at sufficiently high pressure 
the rate constant may be determined by the correlation function at 
short time ($t \alt \eta^{-1}$). 

{}For a short time, the paths which must be considered in 
evaluating a potential $V$ never move very far from the initial position.  
Thus, to a first approximation, we can expand the potential $V$ around 
the average of the initial ($s$) and final ($s'$) positions: 
\begin{equation}
V(x(t)) \simeq V({\bar x}) + [x(t) - {\bar x}]V'({\bar x}) 
+ \frac{1}{2} [x(t) - {\bar x}]^2 V''({\bar x}) , 
\label{pot}
\end{equation}
where ${\bar x}=(s+s')/2$. Here we ignore terms of higher 
order than the second derivative of $V$.  
By virtue of the quadratic form 
of $x(t)$, we can easily find the propagator\cite{path,path2}
\begin{equation}
\langle s' | e^{-i\hat{H}t} | s \rangle = 
\left[ \frac{m\omega}{2\pi i \sin\tau} \right]^{1/2} 
e^{-itV({\bar x})} e^{iS} , 
\label{pr1}
\end{equation}
where 
\begin{equation}
S = \frac{m\omega}{4\sin\tau}(1+\cos\tau)(s-s')^2 - 
\frac{f^2F(\tau)}{2m\omega^3\sin\tau} , 
\label{s}
\end{equation}
with $\tau = \omega t$, 
$f = -V'({\bar x})$, $\omega^2 = V''({\bar x})/m$ and 
$F(\tau) = 2(1-\cos\tau) - \tau\sin\tau$.  Because the potential is 
expanded up to ${\cal O}(V'')$, Eq.(\ref{pr1}) is valid up to 
${\cal O}(f^2)$ or ${\cal O}(\omega^2)$. 

To take the temperature effect into account, it is again necessary to 
introduce complex-times.  Because the correlation function is given by 
(see Eqs.(\ref{gflux}) and (\ref{gcff})) 
\begin{eqnarray}
C^\kappa(r,p;t) &=& \frac{1}{\kappa\beta} 
\int_{(1-\kappa)\beta/2}^{(1+\kappa)\beta/2} 
d\lambda \ 
{\rm tr}[e^{-(\beta-\lambda)\hat{H}}\hat{F}(r)e^{-\lambda\hat{H}}
e^{i\hat{H}t}\hat{F}(p)e^{-i\hat{H}t}] ,  \label{gen1} \\
&=& \frac{1}{\kappa\beta} 
\int_{-\kappa\beta/2}^{\kappa\beta/2} 
d\lambda \ 
{\rm tr}[ \hat{F}(r) e^{i\hat{H}(t+i(\beta/2+\lambda))} \hat{F}(p) e^{-i\hat{H}(t-i(\beta/2-\lambda))}] , \label{gen2} \\
&\equiv& \frac{1}{\kappa\beta} \int_{-\kappa\beta/2}^{\kappa\beta/2} 
d\lambda \ C^\kappa(r,p,\lambda;t) , \label{gen3}
\end{eqnarray}
we define new complex-times as $t_\pm(\lambda) = t\pm i(\beta/2\pm\lambda)$. 
Because of very high pressure (or {\em short} complex-time), it is clear 
that the $(rp)$-rate 
constant becomes very small compared with the $(rr)$-rate constant. 
Therefore, we here focus on the $(rr)$-rate constant and do not consider 
the $(rp)$-rate constant.  Then, using the propagator Eq.(\ref{pr1}) 
we obtain 
\begin{equation}
C^\kappa(r,r,\lambda;t) \simeq \frac{1}{4\pi [t_+(\lambda)t_-(\lambda)]^{3/2}} 
\left[ \frac{\beta}{2} - \frac{\beta^3}{24m}V''(r) \right] 
e^{-\beta V(r)} , 
\label{corr2}
\end{equation}
up to ${\cal O}(V'')$.  The correlation  
$C^\kappa(r,r,\lambda;t)$ is completely determined by $V$ and $V''$, 
while the first derivative $V'$ gives a correction to the 
correlation for the $(rp)$-rate constant. 
Because the quantum correction due to the potential does not depend on 
$\lambda$ explicitly (see Eq.(\ref{corr2})), the correlation function 
$C^\kappa(r,r;t)$ can be 
described in terms of that for the free particle case. 
It is then written best in the form 
\begin{equation}
C^\kappa(r,r;t) = \exp\left[-\beta V(r) - \frac{\beta^2}{12m}V''(r)
\right] \times C_0^\kappa(r,r;t) .  
\label{rr3}
\end{equation}

In a typical chemical reaction, the interatomic distance and forces 
usually range over one or two angstroms, and hence one can expect 
that the change of the potential is small  
while the motion of the order of $1${\AA} has been 
achieved.  Because, in the ideal gas model, the root-mean-square speed of a 
molecule is usually estimated as $v \sim 1.6 
\sqrt{T/A_m}\times 10^{2}$ (m/s), the particle moves by as much as 
$\Delta x \sim 3.5 (T/P\sqrt{A_m}) \times 10^{-4}$ (m) for 
the short period of $\eta^{-1}$.  Thus, if $\Delta x \alt 1$ {\AA}, the 
expansion of the potential Eq.(\ref{pot}) is justified and we can use 
the correlation function Eq.(\ref{rr3}). If we assume a typical 
mass of $A_m \sim 20$ and $T \sim 300$ K, we find the condition of  
$P \agt 0.24$ GPa. 

\begin{figure*}
\includegraphics[height=11cm]{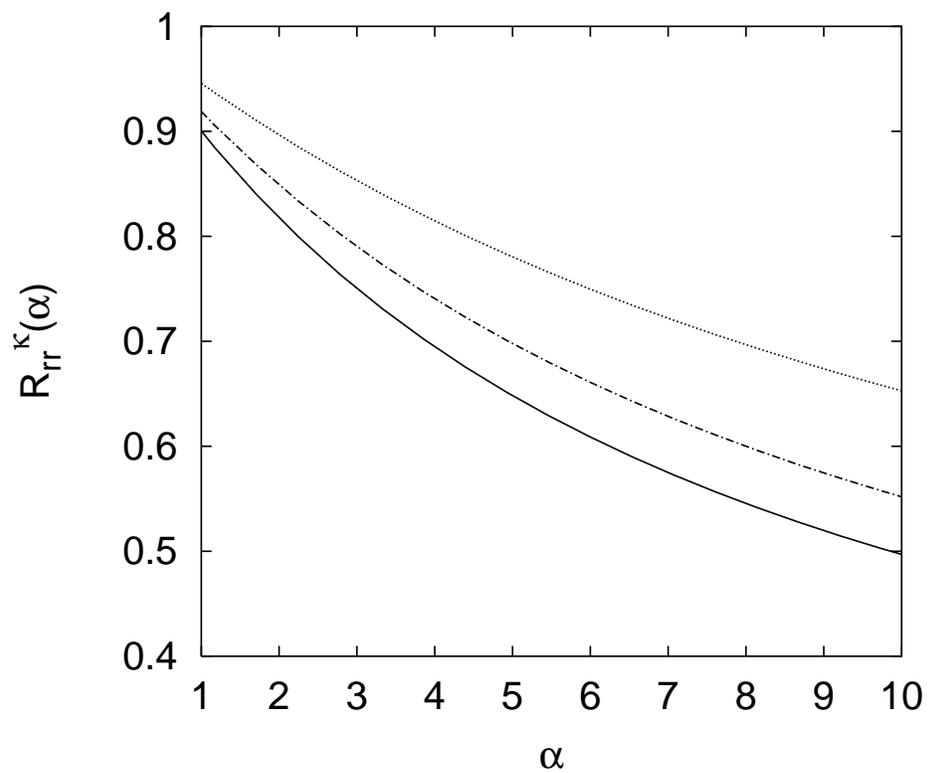}%
\caption{\label{f:hp} Ratio of the $(rr)$-rate constants at 
very high pressure. 
The solid (dot-dashed) [dotted] curve is for $\kappa = 0 (0.5) [0.75]$.}
\end{figure*}

Since the correction due to the potential in Eq.(\ref{rr3}) is 
common to the correlation function for arbitrary $\kappa$, 
the ratio of the $(rr)$-rate constant for arbitrary $\kappa$ to 
that in the Yamamoto approach ($\kappa=1$) is {\em again} given by 
the ratio for the free particle case: 
$R_{rr}^\kappa(\alpha) = k_{rr}^\kappa/k_{rr}^{\kappa=1}$. 
(Note that $R_{rec}^{\kappa} \simeq R_{rr}^{\kappa}$ because 
at high pressure the $(rp)$-rate constant is negligibly small.) 
In Fig.\ref{f:hp}, we show the ratio 
$R_{rr}^\kappa$ in the region of $1 \leq \alpha \leq 10$ 
(cf. Fig.\ref{f:Rrr}). At large $\alpha$ the ratio is reduced 
considerably, which implies that at very high pressure 
the $(rr)$-rate constant for $\kappa \ne 1$ is much smaller than 
that in the Yamamoto approach. 

At high $P$, the rate constant for $\kappa \ne 1$ may be estimated by 
\begin{equation}
Q_0(T) k_{rr}^{\kappa \ne 1} \simeq \frac{Z}{\beta^2} \times 
S_{rr}^{\kappa\ne 1}(0) 
\int_0^\infty dx \ e^{-\alpha x} = 
\frac{2Z}{\alpha\beta\sqrt{1-\kappa^2}} ,   \label{ap1} 
\end{equation}
where $Z$ is a (dimensionless) constant which involves the correction 
due to the potential.  
By contrast, in the case of $\kappa =1$ the rate constant may be given 
by 
\begin{equation}
Q_0(T) k_{rr}^{\kappa = 1} \simeq \frac{Z}{\beta^2} \times 
[\sqrt{x} S_{rr}^{\kappa=1}(x)]_{x \to 0}  
\int_0^\infty dx \ \frac{e^{-\alpha x}}{\sqrt{x}} = 
\frac{Z}{\beta} \sqrt{\frac{\pi}{2\alpha}} .   \label{ap2} 
\end{equation}
The ratio at high pressure is thus given as 
\begin{equation}
R_{rr}^{\kappa} = 
\frac{2}{\sqrt{1-\kappa^2}} \sqrt{\frac{2}{\pi\alpha}} ,  \label{rto}
\end{equation}
which approaches zero in the limit $P \to \infty$. At $\alpha=10$ 
(which, for example, corresponds to the case of 
$T \sim 300$ K and $P \sim 15$ GPa), 
Eq.(\ref{rto}) gives $R_{rr}^{\kappa} = 0.5 (0.58) [0.76]$ for 
$\kappa = 0 (0.5) [0.75]$.  The value for $\kappa =0$ very well agrees with 
the numerical result (see Fig.\ref{f:hp}), while the value for $\kappa = 0.75$ 
is a little larger than the numerical one.  The approximation 
Eq.(\ref{ap1}) may not work well for large $\kappa$ because the peak 
of the scaling function at $x=0$ becomes sharp near $\kappa = 1$ 
(see Fig.\ref{f:srr}). 
In the present calculation, we find that the $(rr)$-rate constant in the
Yamamoto approach is considerably different from the result calculated by 
the {\em partial-split} form.  At $\alpha \sim 10$ the rate constant for 
$\kappa = 1$ is about twice as large as that for $\kappa =0$. 

\section{\label{sec:5}Summary and Conclusion}

The exact quantum mechanical expression for thermal reaction rates can be
formulated by the linear response theory,\cite{kubo,kubo2,mori} which
Yamamoto first discussed in the early 60's.\cite{yama} 
Later, Miller et al.\cite{miller,miller2,miller3} have independently proposed 
a more convenient way, i.e., the flux-flux autocorrelation 
function method, to perform numerical computations.  
The Miller approach can provide the exact rate constant in the limit that
the dynamics of the system is extended to $t \to \infty$. 
Using a general form of the Boltzmannized flux operator, 
we have unified the two approaches and 
studied the rate constants for 
thermal exchange and recombination reactions. Because
they are calculated by 
Laplace transforms of the flux-flux correlation functions, the
result depends on how to choose the Boltzmannized flux operator. 

In this paper, we have first considered a solvable model, i.e., the free
particle case, to demonstrate the dependence of the rate constant 
on $\kappa$ intuitively.  To study it, we have introduced a new 
scaling function and investigated its properties in detail. 
As a non-trivial case, we have discussed the $(rr)$-rate 
constant at very high pressure.  Because under such conditions 
the reaction rate is determined by a propagator at short time, 
the ratio of the rate constant in the {\em partial-split} form of 
the Boltzmannized flux operator to that in the Yamamoto approach can be 
described in terms of the scaling function for the free 
particle case.  We have found that the rate 
constant for recombination reaction in the Yamamoto approach is larger than
that in the case of $\kappa \ne 1$.  In particular, 
the $(rr)$-rate constant in the Yamamoto approach is about twice as 
large as that in the Miller approach at $\alpha \sim 10$. 

In conclusion, the {\em partial-split} form of the Boltzmannized flux 
operator is certainly an economical and powerful tool to
perform numerical calculations for thermal rate constants of realistic reactions. However, for the recombination reaction 
it may underestimate the rate constant compared with the result calculated by 
the linear response theory.  The difference could be seen if the 
experiments for recombination reactions could be 
performed under the conditions of very high pressure. 

\bibliography{kdep}
%

%%%%%%%%% Figure captions %%%%%%%%%%%
%\newpage
%
%\begin{center}
%\underline{\large Figure captions}
%\end{center}
%
%\noindent 1. Sketch of a potential surface in one-dimensional
%reaction versus the reaction coordinate $s$. The dividing point is 
%denoted by $r$.
%
%\noindent 2. One-dimensional schematic diagram of the 
%potential for $A+BC \to AB+C$ reaction.
%The compound region ($ABC$) is bounded by the dividing points on
%reactant ($r$) and product ($p$) sides.
%
%\noindent 3. Same as Fig.\protect\ref{f:diag}, but for 
%the free particle case.
%
%\noindent 4. Scaling function $S_{rr}^\kappa(x)$ for the free 
%particle case.
%The dotted (dot-dashed) [solid] curve is for $\kappa = 0 (0.9) [1]$. 
%
%\noindent 5. Scaling function $S_{rp}^\kappa(x,c)$ for 
%the free particle case.
%The dotted (dot-dashed) [solid] curve is for $\kappa = 0 (0.9) [1]$. 
%We take $c = 1.0$. 
%
%\noindent 6. Ratio of the $(rr)$-rate constants in the free particle case. 
%The solid (dot-dashed) [dotted] curve is for $\kappa = 0 (0.5) [0.75]$. 
%
%\noindent 7. Same as Fig.\protect\ref{f:Rrr}, but for 
%the $(rp)$-rate constant. 
%The solid (dot-dashed) [dotted] curve is for $\kappa = 0 (0.5) [0.75]$. 
%We take $c = 20.0$. 
%
%\noindent 8. Ratio of the $(rr)$-rate constants at very high pressure. 
%The solid (dot-dashed) [dotted] curve is for $\kappa = 0 (0.5) [0.75]$. 
%

\newpage

\newpage

\newpage

\newpage

\newpage

\newpage

\newpage

\newpage

\end{document}